\documentclass[twocolumn,pra,superscriptaddress,showpacs,tightenlines]{revtex4}
\usepackage{amssymb}
\usepackage{amsmath}
\usepackage{graphicx}
\usepackage{epsfig}
\usepackage{subfigure}
\usepackage{amsfonts}
\usepackage{CJK}
\usepackage{txfonts}
\usepackage[colorlinks,citecolor=blue, linkcolor=blue,hyperindex,CJKbookmarks,dvipdfm]{hyperref}

\begin{document}

\title{Controlling single-photon transport in waveguides with finite cross-section}
\author{Jin-Feng Huang }
\affiliation{Advanced Science Institute, RIKEN, Wako-shi, Saitama 351-0198 Japan }
\affiliation{State Key Laboratory of Theoretical Physics, Institute of Theoretical
Physics, Chinese Academy of Sciences, and University of the Chinese
Academy of Sciences, Beijing 100190, China}
\author{Tao Shi }
\affiliation{Max-Planck-Institut f\"{u}r Quantenoptik, Hans-Kopfermann-Strasse 1,
Garching, Germany}
\author{C. P. Sun}
\affiliation{Advanced Science Institute, RIKEN, Wako-shi, Saitama 351-0198 Japan }
\affiliation{Beijing Computational Science Research Center, Beijing 100084, China }
\author{Franco Nori }
\affiliation{Advanced Science Institute, RIKEN, Wako-shi, Saitama 351-0198 Japan }
\affiliation{Department of Physics, University of Michigan, Ann Arbor, Michigan
48109-1040, USA}

\date{\today}
\begin{abstract}
We study the transverse-size effect of a quasi-one-dimensional rectangular
waveguide on the single-photon scattering on a two-level system. We
calculate the transmission and reflection coefficients for single
incident photons using the scattering formalism based on the Lippmann-Schwinger
equation. When the transverse size of the waveguide
is larger than a critical size, we find that the transverse mode will be involved
in the single-photon scattering. Including the coupling to a higher
traverse mode, we find that the photon in the lowest channel will
be lost into the other channel, corresponding to the other transverse
modes, when the input energy is larger than the maximum bound-state
energy. Three kinds of resonance phenomena are predicted: single-photon
resonance, photonic Feshbach resonance, and cutoff (minimum) frequency
resonance. At these resonances, the input photon is completely reflected.
\end{abstract}


\maketitle
\narrowtext

\section{Introduction}

Current optical communications use electronic switching and thus
are limited to electronic speeds of a few gigahertz. To reach much higher
speeds, various proposals have been made including optical networks
~\cite{Kimble2008Nat}, as well as using all-optical routers~\cite{Walsworth2008PRL}
and switches~\cite{Lukin2007NatPhy,Zhou2008PRL,Gong2008PRA,Liao2009PRA,Liao2010PRA,Chang2011PRA}.
Also, quantum optical networks were motivated by
quantum information (communication), using elements with quantum
coherence (such as superposition and entanglement) of photons. Thus
the elemental device can be implemented as a generalized cavity QED
system: a photon confined to a one-dimensional (1D) waveguide, and
controlled by a quantum switch, made of a two (or more) energy-level
systems~\cite{Lukin2007NatPhy,Zhou2008PRL,Gong2008PRA,Liao2009PRA,Liao2010PRA,Chang2011PRA,Hu2007,Shi2009,Liao2010B,Wang2012PRA,Fan2005PRL_OL,Fan2007PRL,Law2008PRA,Shi2009PRB,Liao2010B,Xu2010,Roy2010PRB,Shi2011PRA,Zheng2012PRA,Fan2012PRL,Shi2012,JFHuang2012}.

There have been numerous theoretical~\cite{Lukin2007NatPhy,Zhou2008PRL,Gong2008PRA,Liao2009PRA,Liao2010PRA,Chang2011PRA,Yamamoto1998PRL } and experimental~\cite{Sandoghdar2009Nat,Tsai2010Science}
studies for such a quantum switch, which could be realized in various
physical systems, e.g., a transmission line~\cite{Nori2006PRA,You2003PRB,Mooij2004Nat,Schoelkopf2004Nat,Liao2009PRA}
coupled to a charge qubit~\cite{Simmonds2007Nat,You2005phyTod,Shumeik 2005,YouNat2011,NoriRPP2011} and
a defect cavity waveguide coupled to a quantum dot~\cite{Hughes2004,Hughes2004PRB,Vuckovic2007Nature}.
Most theoretical studies on these systems are excessively idealized,
because the experimental system is never one dimensional.

In order to consider more realistic systems, here we study the finite
cross-section effect of the waveguide on the single-photon transport
controlled by a two-level system (TLS). We consider the waveguide
as a quasi-1D system with a rectangular cross-section. It is well
known that if a photon could be perfectly transported in a quasi-1D
waveguide, its frequency must be larger than the cutoff frequency
of a certain transverse mode. Moreover, to avoid the loss of the photon
incident in the lowest transverse mode due to scattering into other
modes, people need to make the cross section of the waveguide as small
as possible. However, the cross section of realistic waveguides cannot
be infinitely small, and a waveguide with a finite cross section would
allow the photon transit from one transverse mode to another. Furthermore,
if the incident photon frequency is far from the cutoff frequency,
such as x ray~\cite{Rohlsberger2010Sci,Rohlsberger2012Nat}, then
the different transverse modes would be so close that the incident
photon would be inevitably coupled to higher transverse modes. This consideration
motivates us to study the incident photon transport in one mode while
coupled to another (higher) mode.

We solve the Lippmann-Schwinger equation for calculating the reflection
and transmission coefficients of a single photon scattered by a TLS.
Since the exact dispersion relation of a photon in a waveguide with
finite cross section is more like a quadratic one near the cutoff
frequency, quite different from the linear regime, we approximate
the exact dispersion relation by a quadratic function of the wave vector
of the photon by expanding it to second order in the wave vector.
In such a quadratic waveguide, we find that there is a bound state
and two quasibound states for each scattering channel defined by
a certain transverse mode. We note that this bound state does not
exist in the usual linear waveguides.

There are three kinds of resonance phenomena, which correspond to
the complete reflection of the photon incident in a given channel.
One occurs at the single-photon resonance, namely the incident photon
energy is resonant with the TLS without coupling to the higher transverse
mode. Once the incident photon couples to the higher transverse mode,
this resonance phenomenon is replaced by a photonic Feshbach resonance,
namely a complete reflection occurs when the incident energy of the
photon equals the bound-state energy of the higher transverse mode.
The third type of resonance always occurs at the minimum frequency
of the quadratic waveguide, whether or not the singe photon is coupled
to a higher transverse mode. This resonance phenomenon is called cutoff-frequency
resonance. We also notice that the transverse mode will lead to an
incident photon loss as a result of scattering into other higher channels.
We also compare in detail the results obtained by the linear and
quadratic dispersion relations, respectively.

This paper is organized as follows. In Sec.~II, we describe the system
and the effective Hamiltonian, including two transverse modes. We also derive the second-order dispersion relation. Then, we calculate the single-photon transport in the higher transverse
mode without coupling to the incident mode in Sec.~III. We find a
bound state and two quasibound states~\cite{Petrosky2007PRL,Petrosky2008PTP,Gong2008PRA}
by utilizing the quadratic dispersion relation. In Sec.~IV, we obtain
the single-photon reflection and transmission coefficients with coupling
to the higher transverse mode through the Lippmann-Schwinger equation.
The transverse effect in both linear and quadratic waveguides are
discussed in Sec.~V. Finally, we present our conclusions in Sec.~VI.

\section{Model }

\begin{figure}
\includegraphics[bb=0bp 0bp 527bp 328bp,clip,scale=0.45]{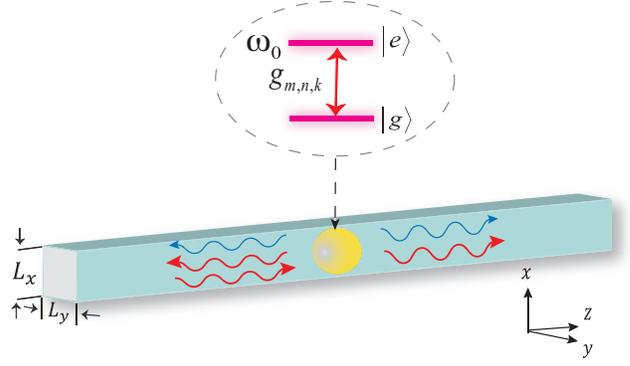}\caption{\label{fig:setup}(Color online) Schematic diagram for single-photon
transport in a quasi-one-dimensional waveguide coupled to a TLS with
transition frequency $\omega_{0}$. The cross-section size of the
waveguide is $L_{x}$ ($L_{y}$) along the $x$ ($y$) direction.}
\end{figure}

The setup under consideration is a waveguide-QED system (see Fig.~\ref{fig:setup})
consisting of a quasi-1D rectangular waveguide with inner dimensions
$L_{x}$ and $L_{y}$ and a two-level atom. The waveguide supports
quantum fields of transverse electric waves $\mbox{T}\mbox{E}_{mn}$,
which are described by the annihilation (creation) operators $a_{m,n,k}^{(\dag)}$.
Here the natural numbers $m$ and $n$ are, respectively, the transverse
quantum numbers in the $x$ and $y$ directions, while the continuous
variable $k$ denotes the wavevector along the $z$ axis. The eigenmode
function of the electric fields in the waveguide can be expressed
as~\cite{Raudorf1978}
\begin{eqnarray}
\widetilde{u}_{m,n,k}^{\left(x\right)}\left(\textbf{r}\right) & = & -i\varepsilon_{k}\frac{2n\pi}{k_{\textrm{\mbox{cut}}}L_{y}}\cos\left(\frac{m\pi}{L_{x}}x\right)\sin\left(\frac{n\pi}{L_{y}}y\right)e^{ikz},\nonumber \\
\widetilde{u}_{m,n,k}^{\left(y\right)}\left(\textbf{r}\right) & = & i\varepsilon_{k}\frac{2m\pi}{k_{\textrm{\mbox{cut}}}L_{x}}\sin\left(\frac{m\pi}{L_{x}}x\right)\cos\left(\frac{n\pi}{L_{y}}y\right)e^{ikz},
\end{eqnarray}
where we introduce the cutoff wavenumber
\begin{eqnarray}
\quad k_{\textrm{\mbox{cut}}}=\sqrt{\left(m\pi/L_{x}\right)^{2}+\left(n\pi/L_{y}\right)^{2}},
\end{eqnarray}
and the electric field per photon $\varepsilon_{k}=\sqrt{\hbar\omega_{m,n,k}/(2\epsilon_{0}V_{k})}$,
with frequency
\begin{eqnarray}
\omega_{m,n,k} & = & c\sqrt{\left(m\pi/L_{x}\right)^{2}+\left(n\pi/L_{y}\right)^{2}+k^{2}},\label{eq:wmn}
\end{eqnarray}
and the effective volume $V_{k}=L_{x}L_{y}2\pi/|k|$ of a segment (with
length $2\pi/|k|$) of the waveguide. The parameter $\epsilon_{0}$
is the vacuum permittivity and $c$ is the speed of light in vacuum.

When a two-level atom is placed in the waveguide, it will couple to
these quantum fields via the dipole interaction. Denoting the ground
and excited states of the atom as $\left|g\right\rangle $ (with energy
$0$) and $\left|e\right\rangle $ (with energy $\omega_{0}$), we
can define the atomic transition operators as $\sigma_{+}=\left|e\right\rangle \left\langle g\right|$
and $\sigma_{-}=\left|g\right\rangle \left\langle e\right|$, and
then the Hamiltonian (with $\hbar=1$) of the waveguide-QED system
reads
\begin{eqnarray}
H & = & \omega_{0}\left|e\right\rangle \left\langle e\right|+\int_{-\infty}^{+\infty}dk\sum_{m,n}\omega_{m,n,k}a_{m,n,k}^{\dagger}a_{m,n,k}\nonumber \\
 &  & +\int_{-\infty}^{+\infty}dk\sum_{m,n}(g_{m,n,k}\sigma_{+}a_{m,n,k}+\mbox{H.c.}).\label{eq:H}
\end{eqnarray}
Here, the coupling strength is $g_{m,n,k}=-d_{e,g}^{\left(x\right)}\widetilde{u}_{m,n,k}^{\left(x\right)}\left(\textbf{r}_{0}\right)-d_{e,g}^{\left(y\right)}\widetilde{u}_{m,n,k}^{\left(y\right)}\left(\textbf{r}_{0}\right)$.

Keeping the coupling between photons and atoms $g_{m,n,k}$ nonzero
requires $mn\neq0$. If $m=0$, namely the coupling along $y$ direction
is zero, then the transverse mode quantum number $n$ should be nonzero,
namely $n=1,2,3, \ldots$; otherwise, if $n=0$,
namely the coupling along $x$ direction is zero, then the transverse
mode quantum number $m$ should be nonzero, namely $m=1,2,3, \ldots$.
When the transverse sizes satisfy $L_{x}=L_{y}$, then the modes
$\mathrm{\textrm{TE}}_{10}$ and $\mathrm{\textrm{TE}}_{01}$, bearing
the same cutoff frequencies, are degenerate. To mainly show our idea,
namely the effect induced by transverse size of the waveguide, we
will choose two modes with different cutoff frequencies. The relation
$\omega_{m,n}^{\textrm{\mbox{cut}}}\equiv ck_{\textrm{\mbox{cut}}}$
gives the exact cutoff frequency $\omega_{m,n}^{\textrm{\mbox{cut}}}$
for the transverse mode $(m,n)$.

As a result of $g_{0,0,k}=0$, we do not consider the $\mbox{T}\mbox{E}_{00}$
mode with $\omega_{0,0,k}=c\left|k\right|$. To reduce the energy
distribution in the transverse mode of the transport photon, we assume
that the photons are in the lowest transverse mode $\mbox{T}\mbox{E}_{01}$,
which is the main transport channel we will consider here. However,
the transverse mode $\mbox{T}\mbox{E}_{11}$ with a little higher energy is very close to the
lowest transverse mode $\mbox{T}\mbox{E}_{01}$ for a finite cross
section of the waveguide, while other transverse modes are far away
from $\mbox{T}\mbox{E}_{01}$. Therefore, the finite cross-section
effect of the quasi-1D waveguide on photon transport can be mainly
characterized by the two transverse modes $\mbox{T}\mbox{E}_{01}$
and $\mbox{T}\mbox{E}_{11}$. Then the Hamiltonian~(\ref{eq:H}) reduces
to
\begin{equation}
H=H_{0}+V\label{H}
\end{equation}
with the free Hamiltonian $H_{0}$ of the photon and the two-level
atom
\begin{eqnarray}
H_{0} & = & H_{w}+\omega_{0}\left|e\right\rangle \left\langle e\right|,
\end{eqnarray}
where
\begin{equation}
H_{w}=\int_{-\infty}^{+\infty}dk(\omega_{a,k}a_{k}^{\dagger}a_{k}+\omega_{b,k}b_{k}^{\dagger}b_{k}),
\end{equation}
and the interaction Hamiltonian $V$ between the photon and the atom
\begin{eqnarray}
V=\int_{-\infty}^{+\infty}dk\sigma_{+}\left(g_{1k}a_{k}+g_{2k}b_{k}\right)+\mbox{H.c.}
\end{eqnarray}
by defining $\mbox{T}\mbox{E}_{01}$ as the $a$ mode, and $\mbox{T}\mbox{E}_{11}$
as the $b$ mode, that is
\begin{eqnarray}
a_{k} & \equiv & a_{0,1,k},\quad b_{k}\equiv a_{1,1,k},
\end{eqnarray}
\begin{eqnarray}
g_{0,1,k} & = & 2id_{e,g}^{\left(x\right)}\sqrt{\frac{\hbar\omega_{ak}}{2\epsilon_{0}V_{k}}}\sin\left(\frac{\pi}{L_{x}}y_{0}\right)\nonumber\\
 & \equiv & g_{1k},\label{eq:g1k}
\end{eqnarray}
\begin{eqnarray}
g_{1,1,k} & = & id_{e,g}^{\left(x\right)}\sqrt{\frac{\hbar\omega_{bk}}{\epsilon_{0}V_{k}}}\cos\left(\frac{\pi}{L_{x}}x_{0}\right)\sin\left(\frac{\pi}{L_{x}}y_{0}\right)\nonumber\\
 & \equiv & g_{2k}, \label{eq:g2k}
\end{eqnarray}
and
\begin{eqnarray}
\omega_{a,k} & \equiv & \omega_{0,1,k},\quad\omega_{b,k}\equiv\omega_{1,1,k},
\end{eqnarray}
\begin{eqnarray}
\omega_{a}^{\texttt{cut}} & \equiv & \omega_{0,1}^{\texttt{cut}},\quad\omega_{b}^{\texttt{cut}}\equiv\omega_{1,1}^{\texttt{cut}}.
\end{eqnarray}

In many works related to 1D waveguides, the dispersion relation of
the photon is approximated up to the first order of the photon wave vector
~\cite{Fan2007PRL,Law2008PRA,Roy2010PRB,Zheng2012PRA,Shi2011PRA,Shi2012,JFHuang2012,Fan2012PRL,Liao2010B,Fan2005PRL_OL}.
However, the exact dispersion relation~(\ref{eq:wmn}) near the cutoff
frequency is more like a quadratic one, so we expand the frequency
$\omega_{a,k}$ around $\left(k_{0},\omega_{0}\right)$ with $\omega_{0}=\omega_{a,k_{0}}=c\sqrt{\pi^{2}L_{y}^{-2}+k_{0}^{2}}$, and $\omega_{b,k}$ around $(k_{0}^{'},\omega_{0})$ with
$\omega_{0}=\omega_{b,k_{0}^{'}}=c\sqrt{\pi^{2}L_{x}^{-2}+\pi^{2}L_{y}^{-2}+k_{0}^{'2}}$, up to second order in $k$. After introducing $p=k-k_{0}$ (for $\omega_{a,k}$),
and $p=k-k_{0}^{'}$ (for $\omega_{b,k}$), the two dispersion relations
can be rewritten as
\begin{eqnarray}
\omega_{s,p} & \simeq & \omega_{0}+v_{s1}p+v_{s2}p^{2}\quad\left(s=a,b\right),\label{eq:wap}
\end{eqnarray}
with the first- and the second-order coefficients given by
\begin{eqnarray}
v_{a1} & = & c\delta/\omega_{0},\quad v_{a2}=\frac{\omega_{0}v_{a1}^{2}}{2\delta^{2}}-\frac{v_{a1}}{2\omega_{0}},\\
v_{b1} & = & \left|v_{a1}\right|\sqrt{2\delta^{2}-\omega_{0}^{2}}/\delta,\quad v_{b2}=2v_{a2}.
\end{eqnarray}

Here, we have introduced $\omega\equiv c\pi/L_{x}$ and $\delta=\pm\sqrt{\omega_{0}^{2}-\omega^{2}}$,
which is proportional to the size $L_{x}$ of the cross section. The
$\pm$ sign represents the sign of $k_{0}$ ($k_{0}^{'}$). The approximated
quadratic dispersion relation (\ref{eq:wap}) shifts the cutoff frequency
from $\omega_{s}^{\textrm{\mbox{cut}}}$ (exact) to $\omega_{s}^{\textrm{\mbox{min}}}=\left(4v_{s2}\omega_{0}-v_{s1}^{2}\right)/\left(4v_{s2}\right)$
(approximated). Here $s=a,\mbox{ }b$.

We assume the photons are entering from the left end of the waveguide
in the $a$ mode; thus for the right-moving photons, $k_{0},\mbox{ \ensuremath{k_{0}^{'}}}>0$,
and $\delta$ takes the $"+"$ sign, while for the left-moving photons,
$k_{0},\mbox{ \ensuremath{k_{0}^{'}}}<0$, and $\delta$ takes the
$"-"$ sign. Therefore, the dispersion relations~(\ref{eq:wap}) can
be rewritten as
\begin{eqnarray}
\omega_{s,k} & \simeq & \begin{cases}
\omega_{0}+\left|v_{s1}\right|k+v_{s2}k^{2}, & k_{0},\mbox{ \ensuremath{k_{0}^{'}}}>0\\
\omega_{0}-\left|v_{s1}\right|k+v_{s2}k^{2}, & k_{0},\mbox{ \ensuremath{k_{0}^{'}}}<0
\end{cases}
 \end{eqnarray}
for $s=a,b$. We note that the terms in the dispersion relation~(\ref{eq:wap})
that depend on the photon wave vector $p$ describe the frequency
detuning of the photon from the atom. Later on, we will use the dispersion
relations~(\ref{eq:wap}) in our derivations.

\section{Scattering and bound states in the single $b$ mode}

Since the photon scattering process in the so-called $b$ mode may
contribute to the photon transport in the $a$ mode, we first consider
the photon scattering in a single $b$ mode. We inject the photon
in the $b$ mode with the atom only coupled to the transverse-mode
$b$ mode ($g_{1}=0$). By employing the Lippmann-Schwinger equation,
we calculate the scattering state of the photon in the $b$ mode.
The bound state is also obtained by the poles of the $T$ matrix~\cite{Taylor1972}.

Under the above consideration, the Hamiltonian is directly obtained
by setting $g_{1}=0$ and $\omega_{a,k}=0$ in the Hamiltonian~(\ref{H})
$H^{b}=H_{0}^{b}+V^{b}$, which includes the free Hamiltonian
\begin{eqnarray}
H_{0}^{b} & = & H_{w}^{b}+\omega_{0}\left|e\right\rangle \left\langle e\right|
\end{eqnarray}
with $H_{w}^{b}=\int_{-\infty}^{+\infty}dk\;\omega_{b,k}\; b_{k}^{\dagger}b_{k}$,
and the interaction part
\begin{equation}
V^{b}=\int_{-\infty}^{+\infty}dk(g_{2k}\sigma_{+}b_{k}+\mbox{H.c.}).
\end{equation}
We assume the single photon is initially input from the left end of
the waveguide in the $b$ mode $b_{k}^{\dagger}\left|\emptyset\right\rangle $
with energy $\omega_{b,k}$, while the atom is in the ground state
$\left|g\right\rangle $, then the scattering state is given by the
Lippmann-Schwinger equation~\cite{Taylor1972,PZhang2012PRA}
\begin{equation}
\left|\varphi_{bk}^{\left(+\right)}\right\rangle =b_{k}^{\dagger}\left|\emptyset\right\rangle \left|g\right\rangle +\frac{1}{\omega_{b,k}+i0^{+}-H_{0}^{b}}V^{b}\left|\varphi_{bk}^{\left(+\right)}\right\rangle .\label{eq:psi1-2}
\end{equation}
Here, the input state $b_{k}^{\dagger}\left|\emptyset\right\rangle \left|g\right\rangle $
is the eigenstate of the free Hamiltonian $H_{0}^{b}$ with eigenenergy
$\omega_{b,k}$,
\begin{equation}
H_{0}^{b}b_{k}^{\dagger}\left|\emptyset\right\rangle \left|g\right\rangle =\omega_{b,k}b_{k}^{\dagger}\left|\emptyset\right\rangle \left|g\right\rangle,
\end{equation}
and $\left|\varphi_{bk}^{\left(+\right)}\right\rangle $ is the eigenstate
of the total Hamiltonian $H^{b}$ with the same eigenenergy $\omega_{b,k}$.

We assume that the solution of the scattering state $\left|\varphi_{bk}^{\left(+\right)}\right\rangle $
is in the form
\begin{equation}
\left|\varphi_{bk}^{\left(+\right)}\right\rangle =\left|\phi_{b,k}\right\rangle \left|g\right\rangle +\beta_{b,k}\left|\emptyset\right\rangle \left|e\right\rangle .\label{eq:psi2-2}
\end{equation}
Here, $\left|\phi_{b,k}\right\rangle $ is the single-photon state
after being scattered, and $\beta_{b,k}$ is the probability amplitude
for the atom to be in its excited state. Substituting this solution
into Eq.~(\ref{eq:psi1-2}), the scattering state is obtained,
\begin{eqnarray}
\left|\varphi_{bk}^{\left(+\right)}\right\rangle  & = & b_{k}^{\dagger}\left|\emptyset\right\rangle \left|g\right\rangle +\beta_{b,k}\left|\emptyset\right\rangle \left|e\right\rangle \nonumber \\
 &  & +G_{bw}^{0}\left(\omega_{b,k}+i0^{+}\right)\beta_{bk}\int_{-\infty}^{+\infty}dk'g_{2k'}^{*}b_{k'}^{\dagger}\left|\emptyset\right\rangle \left|g\right\rangle, \nonumber \\
\end{eqnarray}
where $G_{bw}^{0}\left(z\right)=\left(z-H_{w}^{b}\right)^{-1}$ is
the free Green operator for the $b$-mode photon, and
\begin{eqnarray}
\beta_{bk} & = & \frac{g_{2k}}{\omega_{b,k}+i0^{+}-\omega_{0}-\Sigma_{b}\left(\omega_{b,k}\right)},
\end{eqnarray}
with the self-energy defined by
\begin{eqnarray}
\Sigma_{b}\left(E\right) & \equiv &\int_{-\infty}^{+\infty}dk\frac{\left|g_{2k}\right|^{2}}{E+i0^{+}-\omega_{b,k}} \label{eq:Sel-energyb-1}\\
& \approx& -\frac{i\gamma_{b}v_{b1}}{\sqrt{v_{b1}^{2}+4v{}_{b2}\left(E-\omega{}_{0}\right)}}.\label{eq:sel-enegyb-1}
\end{eqnarray}
Here, if we directly substitute the exact coupling expression~(\ref{eq:g2k}) into Eq.~(\ref{eq:Sel-energyb-1}), the divergence of self-energy $\Sigma_{b}$ occures. In obtaining the result~(\ref{eq:sel-enegyb-1}), we have assumed $g_{2k}$ to be independent of
$k$, namely $g_{2k}=g_{2}$. This assumption is equivalent to the Markov approximation~\cite{Gardiner1985}.

It follows from Eq.~(\ref{eq:sel-enegyb-1}) that, when $E\geq\left(4v{}_{b2}\right)^{-1}\left(4v{}_{b2}\omega{}_{0}-v_{b1}^{2}\right)=\omega_{bk}^{\textrm{\mbox{min }}}$,
$\Sigma_{b}\left(E\right)$ is purely imaginary, while $E<\omega_{bk}^{\textrm{\mbox{min}}}$,
$\Sigma_{b}\left(E\right)$ is real. Using the scattering state, the
$T$-matrix elements are given by
\begin{eqnarray}
t_{k'k}\left(\omega_{b,k}\right) & = & \left\langle g\right|\left\langle \emptyset\right|b_{k'}V^{b}|\varphi_{bk}^{(+)}\rangle=\beta_{bk}g_{2k'}^{*}.
\end{eqnarray}

The bound state can be obtained by solving the transcendental equation
$\left[t_{k'k}\left(E_{bs}\right)\right]^{-1}=0$. We directly obtain
the bound-state-energy transcendental equation
\begin{eqnarray}
E_{bs} & = & \omega_{0}-\frac{i\gamma_{b}v_{b1}}{\sqrt{v_{b1}^{2}+4v{}_{b2}\left(E_{bs}-\omega{}_{0}\right)}}\label{eq:Es}
\end{eqnarray}
by using the result~(\ref{eq:sel-enegyb-1}). Here we have defined
the decay rate for the atom induced by the $b$-mode $\gamma_{b}=2\pi\left|g_{2}\right|^{2}/v_{b1}$.
Later on, we use $\gamma_{b}$ to denote the coupling strength $g_{2}$.

It follows from this result~(\ref{eq:Es}) that if $v_{b2}\rightarrow0$,
which corresponds to a linear waveguide, the bound-state-energy solution
is
\begin{eqnarray}
E_{bs} & = & \omega_{0}-i\left|\gamma_{b}\right|.
\end{eqnarray}
The fact that there is no real solution means that there is no bound
state in the linear waveguide. However, for the quadratic waveguide,
the transcendental equation~(\ref{eq:Es}) gives one real solution
\begin{eqnarray}
E_{bs} & = & \Delta_{a}^{\textrm{\mbox{F}}}+\omega_{0},\label{eq:bs}
\end{eqnarray}
with
\begin{equation}
\Delta_{a}^{\textrm{\mbox{F}}}\equiv\frac{u_{b}^{2}-u_{b}v_{b1}^{2}+v_{b1}^{4}}{12u_{b}v_{b2}}.
\end{equation}
This real solution denotes a bound state. Two complex solutions
\begin{eqnarray}
 &  & E_{bs}\nonumber \\
 & = & \left[\left(-1+i\sqrt{3}\right)u_{b}-2v_{b1}^{2}+\left(-1-i\sqrt{3}\right)\frac{v_{b1}^{4}}{u_{b}}+24v_{b2}\omega_{0}\right]\nonumber \\
 &  & \times\left(24v_{b2}\right)^{-1}\nonumber \\
\end{eqnarray}
and
\begin{eqnarray}
 &  & E_{bs}\nonumber \\
 & = & \left[\left(-1-i\sqrt{3}\right)u_{b}-2v_{b1}^{2}+\left(-1+i\sqrt{3}\right)\frac{v_{b1}^{4}}{u_{b}}+24v_{b2}\omega_{0}\right]\nonumber \\
 &  & \times\left(24v_{b2}\right)^{-1},\nonumber \\
\end{eqnarray}
correspond to two quasibound states. Each of these is a metastable
state that decays on a very long time scale and appears to be a localized
bound state in real space~\cite{Petrosky2007PRL,Petrosky2008PTP}.
The parameters therein are defined by
\begin{eqnarray}
u_{b}^{3} & = & l_{b},\nonumber \\
l_{b} & = & -v_{b1}^{6}-216v_{b1}^{2}v_{b2}^{2}\gamma_{b}^{2}\nonumber \\
 &  & +12\sqrt{3}\sqrt{v_{b1}^{4}v_{b2}^{2}\gamma_{b}^{2}\left(v_{b1}^{4}+108v_{b2}^{2}\gamma_{b}^{2}\right)}.
\end{eqnarray}
We note that $l_{b}$ is always real, thus there are three values
for $u_{b}=l_{b}^{1/3},l_{b}^{1/3}e^{i\frac{2\pi}{3}},$ and $l_{b}^{1/3}e^{i\frac{4\pi}{3}}$.
However, we can choose $u_{b}$ to be real. Then there is always
a real solution (\ref{eq:bs}) for Eq.~(\ref{eq:Es}), which
describes a bound state with energy (\ref{eq:bs}). In addition, when
the detuning $\Delta_{ak}\equiv\omega_{ak}-\omega_{0}$ satisfies
\begin{equation}
\Delta_{ak}=\Delta_{a}^{\textrm{\mbox{F}}}
\end{equation}
or equivalently,
\begin{equation}
k=\frac{-v_{a1}\pm\sqrt{v_{a1}^{2}+4v_{a2}\Delta_{a}^{\textrm{\mbox{F}}}}}{2v_{a2}}\equiv k_{\textrm{\mbox{F}}},
\end{equation}
the input photon energy $\omega_{ak}$ is resonant with the bound
state in the $b$ mode. This is the Feshbach resonance. Moreover,
\begin{equation}
\lim_{\gamma_{b}\rightarrow0}\Delta_{a}^{\textrm{\mbox{F}}}=-\frac{v_{b1}^{2}}{4v_{b2}}=\Delta_{\textrm{\mbox{max}}}^{\textrm{\mbox{F}}}.
\end{equation}
This maximum value $\Delta_{\textrm{\mbox{max}}}^{\textrm{\mbox{F}}}$
of $\Delta_{a}^{\textrm{\mbox{F}}}$ versus the coupling strength
$\gamma_{b}$ between the transverse mode and the TLS denotes the
maximum value of the bound-state energy in this transverse mode $E_{bs}^{\textrm{\mbox{max}}}=\Delta_{\textrm{\mbox{max}}}^{\textrm{\mbox{F}}}+\omega_{0}$.

\section{Photon transmission and reflection in the $a$ mode while the atom
is coupled to the $b$ mode}

Now we consider the photon injected in the $a$ mode with the atom
coupled to the $a$ and $b$ modes at the same time. We calculate
the scattering state of the $a$-mode photon. Using the Lippmann-Schwinger
equation, the scattering state is
\begin{equation}
\left|\psi_{k}^{\left(+\right)}\right\rangle =a_{k}^{\dagger}\left|\emptyset\right\rangle \left|g\right\rangle +\frac{1}{\omega_{a,k}+i0^{+}-H_{0}}V\left|\psi_{k}^{\left(+\right)}\right\rangle. \label{eq:psi1}
\end{equation}

By a similar procedure to the last section, the scattering state is
obtained as
\begin{eqnarray}
 &  & \left|\psi_{k}^{\left(+\right)}\right\rangle \nonumber \\
 & = & a_{k}^{\dagger}\left|\emptyset\right\rangle \left|g\right\rangle +\beta_{k}\left|\emptyset\right\rangle \left|e\right\rangle \nonumber \\
 &  & +G_{w}^{0}\left(\omega_{a,k}+i0^{+}\right)\beta_{k}\int_{-\infty}^{+\infty}dk'\left(g_{1k'}^{*}a_{k'}^{\dagger}+g_{2k'}^{*}b_{k'}^{\dagger}\right)\left|\emptyset\right\rangle \left|g\right\rangle ,\nonumber \\
\label{eq:psi2-1}
\end{eqnarray}
where the similar free Green operator is $G_{w}^{0}\left(z\right)=\left(z-H_{w}\right)^{-1}$,
and the excited probability amplitude of the atom is
\begin{eqnarray}
\beta_{k} & = & \frac{g_{1k}}{\omega_{a,k}+i0^{+}-\omega_{0}-\Sigma_{a}\left(\omega_{a,k}\right)-\Sigma_{b}\left(\omega_{a,k}\right)}\label{eq:beta}
\end{eqnarray}
with the self-energy for the $a$ mode defined by
\begin{eqnarray}
\Sigma_{a}\left(E\right)&\equiv&\int_{-\infty}^{+\infty}dk\frac{\left|g_{1k}\right|^{2}}{E+i0^{+}-\omega_{a,k}} \label{eq:Sel-energya-1}\\
 & \simeq & -\frac{i\gamma_{a}v_{a1}}{\sqrt{v_{a1}^{2}+4v{}_{a2}\left(E-\omega{}_{0}\right)}},\label{eq:sel-enegya-1}
\end{eqnarray}
and $\Sigma_{b}\left(E\right)$ defined by Eq.~(\ref{eq:Sel-energyb-1}). Similarly, if we directly substitute the exact coupling expression~(\ref{eq:g1k}) into Eq.~(\ref{eq:Sel-energya-1}), the divergence of self-energy $\Sigma_{a}$ also occurs. In obtaining the result~(\ref{eq:sel-enegya-1}), we have also assumed $g_{1k}$ to be independent of
$k$, namely $g_{1k}=g_{1}$.
Here, the decay rate induced by the $a$ mode $\gamma_{a}=2\pi\left|g_{1}\right|^{2}/v_{a1}$
is introduced. Later on, we also use $\gamma_{a}$ to denote the coupling
strength $g_{1}$.

By using the scattering state (\ref{eq:psi2-1}), we obtain the matrix
elements of the scattering operator $S$ in $k$ space,
\begin{equation}
S_{k',k}=\delta\left(k'-k\right)-2\pi i\delta\left(\omega_{ak'}-\omega_{ak}\right)t_{k'k}\left(\omega_{a,k}+i0^{+}\right),
\end{equation}
where the $T$-matrix elements are directly obtained as
\begin{eqnarray}
 t_{k'k}\left(\omega_{a,k}+i0^{+}\right)
  =\left\langle g\right|\left\langle \emptyset\right|a_{k'}V\left|\psi_{k}^{\left(+\right)}\right\rangle=\beta_{k}g_{1k'}^{*},
\end{eqnarray}
and the $\delta$ function here is defined as $\delta\left(x\right)=1$
at $x=0$; otherwise, $\delta\left(x\right)=0$.

Through the relation
\begin{eqnarray}
S_{k',k}=r\delta\left(k+k'\right)+t\delta\left(k-k'\right),
\end{eqnarray}
we obtain the reflection amplitude
\begin{equation}
r\left(k\right)=-i\frac{1}{\left\vert 2v_{a2}k+\left|v_{a1}\right|\right\vert }\frac{\gamma_{a}v_{a1}}{\Delta_{ak}-\Sigma_{a}\left(\omega_{a,k}\right)-\Sigma_{b}\left(\omega_{a,k}\right)}\label{eq:rk}
\end{equation}
for the input single photon in the $a$ mode. Note that, in obtaining
the result~(\ref{eq:rk}), we have discarded the term proportional
to $\bar{\delta}\left[\Delta_{ak}-\Sigma_{a}\left(\omega_{a,k}\right)-\Sigma_{b}\left(\omega_{a,k}\right)\right]$
and the principle value label $\mathcal{P}$ when using the formula
$1/\left(x+i0^{+}\right)=\mathcal{P}/x-i\pi\bar{\delta}\left(x\right)$.
Here the Dirac $\delta$ function is defined as $\bar{\delta}\left(x\right)=\infty$
if $x=0$; otherwise, $\bar{\delta}\left(x\right)=0$. This procedure
is reasonable because the definition of detuning $\Delta_{ak}$ already
restricts its regime to $\Delta_{ak}\geq-v_{a1}^{2}/\left(4v{}_{a2}\right)=\Delta_{ak}^{\textrm{\mbox{min}}}$,
which contradicts the basic condition $\Delta_{ak}<-v_{a1}^{2}/\left(4v{}_{a2}\right)$,
under which the $\delta$ term $\bar{\delta}\left[\Delta_{ak}-\Sigma_{a}\left(\omega_{a,k}\right)-\Sigma_{b}\left(\omega_{a,k}\right)\right]$
may contribute.

In terms of the detuning $\Delta_{ak}$, the reflection amplitude
is
\begin{eqnarray}
r\left(\Delta_{ak}\right) & = & -i\frac{1}{\sqrt{v_{a1}^{2}+4v_{a2}\Delta_{ak}}}\nonumber \\
 &  & \times\frac{\gamma_{a}v_{a1}}{\Delta_{ak}-\Sigma_{a}\left(\Delta_{ak}+\omega_{0}\right)-\Sigma_{b}\left(\Delta_{ak}+\omega_{0}\right)}.\nonumber \\
\end{eqnarray}
Then the reflection coefficient $R=\left|r\right|^{2}$ can be directly
obtained.

The transmission amplitude $t$ is directly obtained through $t=1+r$
and the transmitted coefficient is straightforwardly obtained as $T=\left|t\right|^{2}$.
Interestingly, we find three resonance points where the single-photon
transmission amplitude is zero: (1) $t\left(\Delta_{ak}=\Delta_{a}^{\textrm{\mbox{F}}}\right)=0$,
where $\Delta_{ak}=\Delta_{a}^{\textrm{\textrm{\mbox{F}}}}$ means that the input single-photon energy is resonant with the bound-state energy
in the transverse mode $\omega_{ak}=E_{bs}$, namely the photonic
Feshbach resonance; (2) $t\left(\Delta_{ak}=0,\gamma_{b}=0\right)=0$,
or in terms of the wave vector $t\left(k=k_{\textrm{\mbox{res}}},\gamma_{b}=0\right)=0$,
with $k_{\textrm{\mbox{res}}}=0,\mbox{ }-v_{a1}/v_{a2}$ for $v_{a1}>0$.
This resonance is denoted as single-photon resonance. (3) $\underset{\Delta_{ak}\rightarrow\Delta_{ak}^{\textrm{\mbox{min}}}}{\lim}t\left(\Delta_{ak}\right)=0$.
We call this resonance the cutoff (minimum) frequency resonance. Under
these three resonances, the transmission $T=0$.

\begin{figure}
\includegraphics[bb=10bp 0bp 513bp 605bp,clip,width=4.5cm,height=6cm]{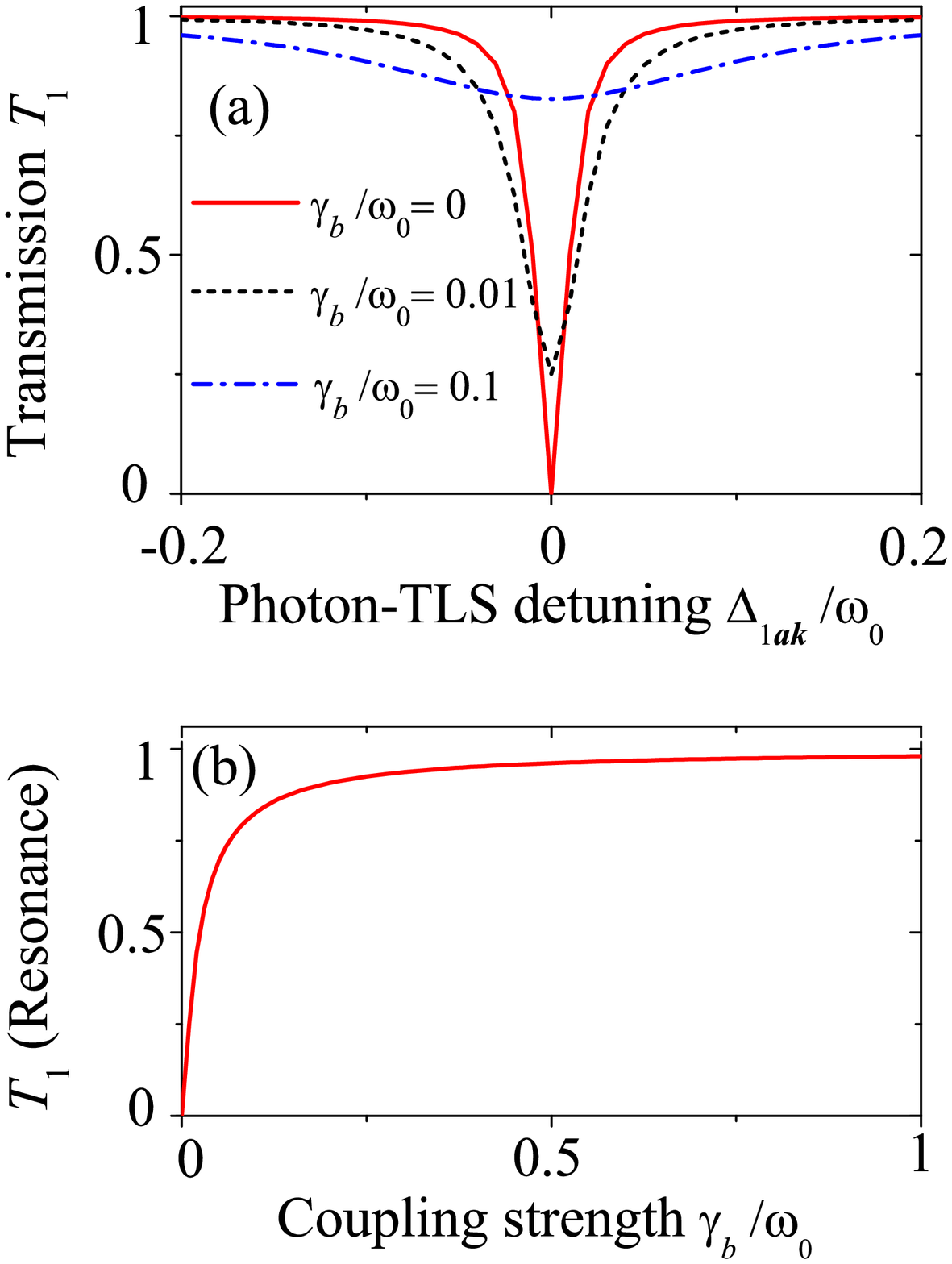}\includegraphics[width=4.5cm,height=6cm]{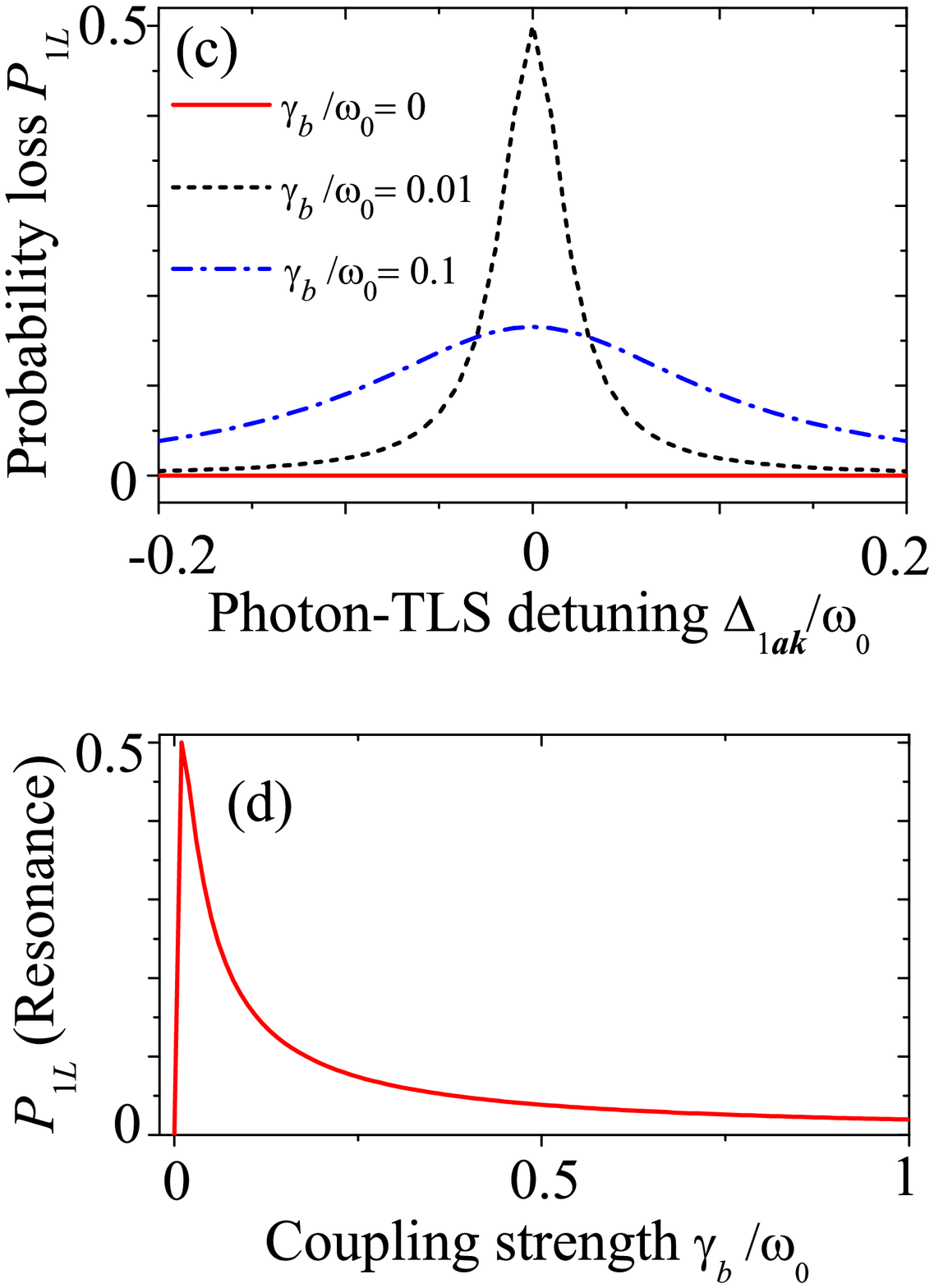}\caption{\label{fig:T1}(Color online) Results for linear waveguides. (a) Transmission
coefficient $T_{1}$ versus detuning $\Delta_{1ak}$ and (b) versus
the coupling strength $\gamma_{b}$ between the transverse mode and
the TLS at the single-photon resonance $\Delta_{1ak}=0$.  (c) The
single-photon loss probability $P_{1L}$ versus detuning $\Delta_{1ak}$
and (d) versus the coupling strength $\gamma_{b}$ between the transverse
mode and the TLS at the single-photon resonance $\Delta_{1ak}=0$.
Other parameters are $\gamma_{a}/\omega_{0}=0.01$, $\delta/\omega_{0}=0.8$,
and $v_{a1}=1$. All the parameters are in units of $\omega_{0}$.}
\end{figure}

For comparison, we also obtain the reflection amplitude
\begin{eqnarray}
r_{1} & = & \frac{-i\gamma_{a}}{\Delta_{ak}+i\left(\gamma_{a}+\gamma_{b}\right)},
\end{eqnarray}
and the transmission amplitude
\begin{equation}
t_{1}=1+r_{1}=\frac{\Delta_{ak}+i\gamma_{b}}{\Delta_{ak}+i\left(\gamma_{a}+\gamma_{b}\right)}
\end{equation}
for linear waveguides and add a subscript "$1$" to denote that this
result only applies to linear waveguides. This result is in agreement
with Refs.~\cite{Fan2005PRL_OL,JFHuang2012} when $\gamma_{b}=0$. Correspondingly,
the reflection and transmission coefficients are
\begin{eqnarray}
R_{1} & = & \left|r_{1}\right|^{2}=\frac{\gamma_{a}^{2}}{\Delta_{1ak}^{2}+\left(\gamma_{a}+\gamma_{b}\right)^{2}}\label{eq:R1}
\end{eqnarray}
and
\begin{eqnarray}
T_{1} & = & \left|t_{1}\right|^{2}=\frac{\Delta_{1ak}^{2}+\gamma_{b}^{2}}{\Delta_{1ak}^{2}+\left(\gamma_{a}+\gamma_{b}\right)^{2}}.\label{eq:T1}
\end{eqnarray}
Here, $\Delta_{1ak}=v_{a1}k$ is the detuning of the single photon
for the $a$ mode in the linear waveguide from the two-level atom.

For linear waveguides, it follows from Eq.~(\ref{eq:R1}) that the
transverse mode will reduce the reflection of the single photon. For
the transmission of the photon, the transverse mode will increase
the transmission of the single photon when $\gamma_{a}\gamma_{b}>2\Delta_{ak}^{2}$;
otherwise, it will decrease its transmission. In addition, as a result
of the transverse mode, i.e., $\gamma_{b}\neq0$, the single-photon
probability is not conserved in its input mode, that is $T_{1}+R_{1}<1$.
The photon is scattered into the transverse mode with probability
\begin{eqnarray}
P_{1L} & \equiv & 1-R_{1}-T_{1}
  =  \frac{2\gamma_{a}\gamma_{b}}{\Delta_{ak}^{2}+\left(\gamma_{a}+\gamma_{b}\right)^{2}}.\label{eq:P1L}
\end{eqnarray}
This probability loss has a Lorentz shape centered at the single-photon
resonance $\Delta_{ak}=0$ with width $\gamma_{a}+\gamma_{b}$. Under
the resonance condition $\Delta_{ak}=0$ and identical coupling to
both scattering mode ($a$ mode) and transverse mode ($b$ mode),
namely, $\gamma_{a}=\gamma_{b}$, the loss probability reaches $P_{1L}=0.5$.
\begin{figure}
\includegraphics[bb=22bp 35bp 317bp 449bp,clip,scale=0.85]{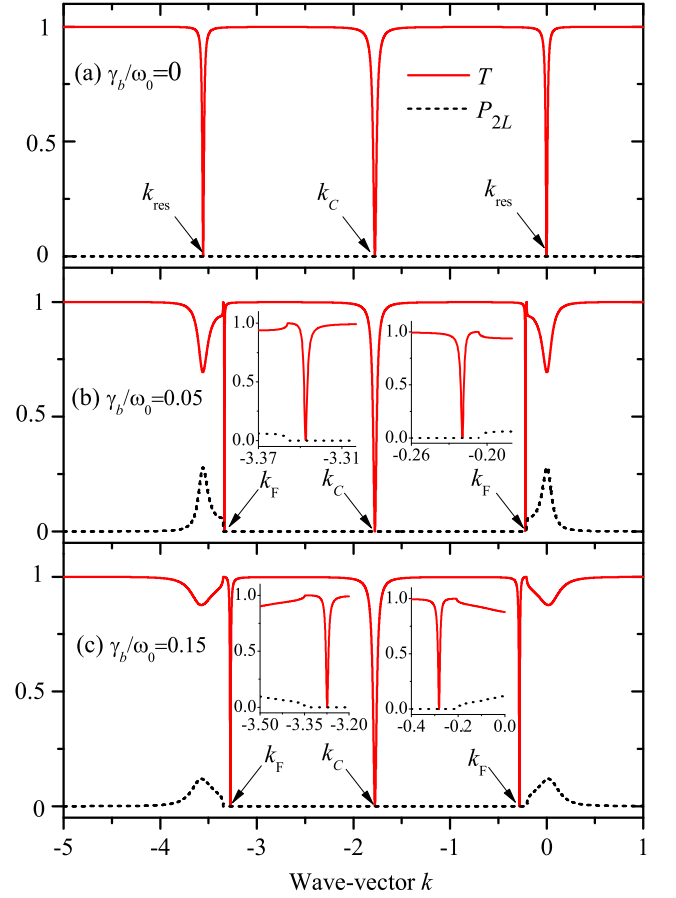}\caption{\label{fig:T}(Color online) Results for quadratic waveguides: transmission
coefficient $T$ and probability loss $P_{2L}$ versus wave vector
$k$ when (a) $\gamma_{b}/\omega_{0}=0$, (b) $\gamma_{b}/\omega_{0}=0.05$,
and (c) $\gamma_{b}/\omega_{0}=0.15$. Other parameters are the same
as in Fig. \ref{fig:T1}.}
\end{figure}

\section{Transverse effect in linear and quadratic waveguides }

\subsection{Transverse effect in linear waveguides }

To show the transverse effect on single-photon transport, we first
consider its effect in linear waveguides. We plot the transmission
coefficient $T_{1}$ and $P_{1L}$ versus detuning $\Delta_{1ak}$
($=k$ with $v_{a1}=1$) under different transverse coupling strengths
$\gamma_{b}/\omega_{0}=0,$ $0.01$, and $0.1$ in Figs.~\ref{fig:T1}(a)
and~\ref{fig:T1}(c), respectively, and versus the coupling strength $\gamma_{b}$
under the single-photon resonance condition $\Delta_{1ak}=0$ in Figs.~\ref{fig:T1}(b) and~\ref{fig:T1}(d), respectively.

It follows from Fig.~\ref{fig:T1}(a) that, at the single-photon resonance
condition, the perfect reflection ($T_{1}=0$) of the single photon
is damaged by the transverse mode, and the width of the transmission
energy band increases as the transverse-mode coupling strength increases.
When the transverse-mode coupling strength is strong enough, the perfect
reflection becomes perfect transmission {[}Fig.~\ref{fig:T1}(b){]}. Furthermore,
the transverse mode forces the single photon to leave the input mode
if the input photon is near resonance with the atom. Especially, exactly
at the single-photon resonance, the loss probability reaches its largest
value. However, this largest value at the single-photon resonance
not always increases as the transverse-mode coupling strength increases,
as shown in Fig.~\ref{fig:T1}(d). It first increases rapidly to $0.5$
at $\gamma_{b}=\gamma_{a}$, then decreases gradually as the transverse-mode coupling strength increases and finally reaches zero when $\gamma_{b}$
is strong enough.
\begin{figure}
\includegraphics[bb=24bp 5bp 450bp 589bp,clip,scale=0.5]{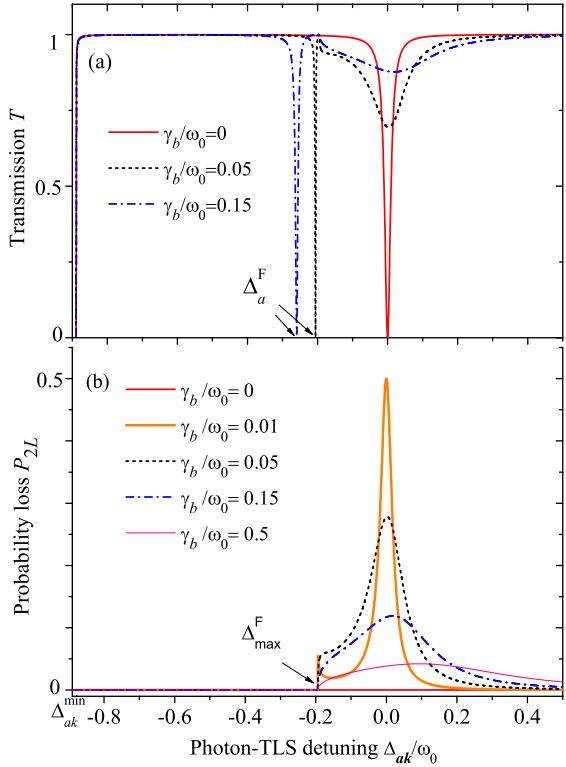}\caption{\label{fig:T_P2L}(Color online) Results for quadratic waveguides.
(a) Transmission coefficient $T$ versus detuning $\Delta_{ak}$ in
quadratic waveguide. (b)The single-photon loss probability $P_{2L}$
versus detuning $\Delta_{ak}$ . Other parameters are the same as
in Fig. \ref{fig:T1}.}
\end{figure}


\subsection{Transverse effect in quadratic waveguides }

Now we illustrate the transverse effect on the single-photon transport
properties in a quadratic waveguide. We plot the transmission coefficient
$T$ versus wave vector $k$ in Fig.~\ref{fig:T} and versus the detuning
$\Delta_{ak}$ in Fig.~\ref{fig:T_P2L}(a), and the loss probability
$P_{2L}=1-T-R$ versus the detuning $\Delta_{ak}$ in Fig.~\ref{fig:T_P2L}(b).

Figure~\ref{fig:T}(a) shows that the single photon is perfectly reflected
at $k=k_{\textrm{\mbox{res}}}$ and $k=k_{C}$; otherwise, it is completely
transmitted without coupling to the transverse mode. However, once
the TLS is coupled to the transverse mode, the original perfect-reflection
points $k=k_{\textrm{\mbox{res}}}$ have been shifted to $k=k_{\textrm{\mbox{F}}}$
with some probability loss at $k=k_{\textrm{\mbox{res}}}$. When increasing
the coupling strength of the transverse mode, the two sides of the
perfect reflection peaks at $k=k_{\textrm{\mbox{F}}}$ move toward
the center peak at $k=k_{C}$, while the probability loss at $k=k_{\textrm{\mbox{F}}}$
is reduced. We also note that the center perfect-reflection peak at
$k=k_{C}$ is not dependent on the transverse-mode coupling; It is
decoupled from the transverse mode. This is because it is only determined
by the minimum detuning $\Delta_{ak}^{\textrm{\mbox{min}}}$ between
the photon and the TLS. Compared with the linear waveguide, this phenomenon
is more robust against the finite cross-section effect of the waveguide.
Also, there are two additional perfect-reflection peaks. Between the peaks
$k_{\textrm{\mbox{F}}}<k<k_{C}$ (or $k_{C}<k<k_{\textrm{\mbox{F}}}$),
there is a perfect transmission band.
\begin{figure}
\includegraphics[bb=24bp 4bp 387bp 310bp,clip,scale=0.65]{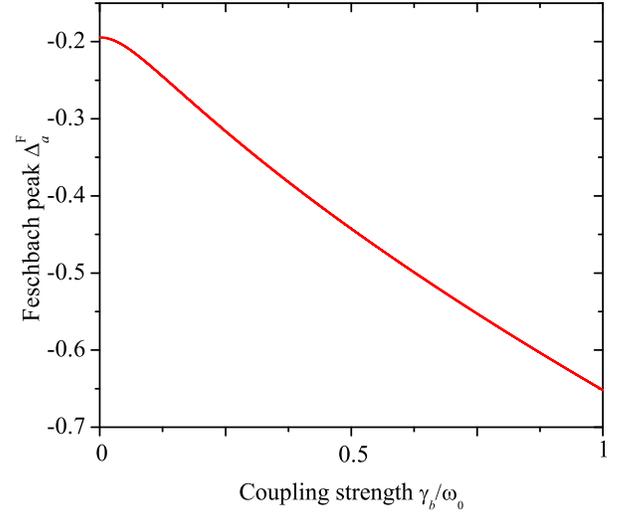}\caption{\label{fig:Detu_Fesh_gamb}Feschbach peak $\Delta_{a}^{\textrm{\mbox{F}}}$
versus the transverse-mode coupling strength $\gamma_{b}$ for waveguides
in the quadratic regime. Other parameters are the same as in Fig.
\ref{fig:T1}.}
\end{figure}

Figure~\ref{fig:T_P2L} shows the single-photon transport properties
in terms of the input energy. Without coupling to the transverse mode,
the single photon is perfectly reflected at $\Delta_{ak}=\Delta_{ak}^{\textrm{\mbox{min}}}$,
$\Delta_{ak}=0$ (single-photon resonance). However, as a result of
the coupling to the transverse mode, the perfect reflection at the
single-photon resonance disappears but it is replaced by another perfect
reflection at $\Delta_{ak}=\Delta_{a}^{\textrm{\mbox{F}}}$, which
denotes that the input single-photon energy is resonant with the bound-state
energy in the transverse mode. This is the photonic Feshbach resonance
~\cite{Feshbach,Xu2010,Kerman1999PR}. Moreover, the position of the
perfect reflection as a result of photonic Feshbach resonance moves
away from the single-photon resonance position. Figure~\ref{fig:T_P2L}(b)
shows that the photon loss probability only occurs in the regime $\Delta_{ak}>\mbox{\ensuremath{\Delta}}_{\textrm{\mbox{max}}}^{\textrm{\mbox{F}}}$.
This is because
\begin{equation}
\underset{\Delta_{ak}\rightarrow\Delta_{\textrm{\mbox{max}}}^{\textrm{\mbox{F}}}}{\lim}P_{2L}=0.
\end{equation}
When $\Delta_{ak}\leq\Delta_{\textrm{\mbox{max}}}^{\textrm{\mbox{F}}}$,
the loss probability becomes zero $P_{2L}=0$. Therefore, when the
single-photon input energy satisfies $\Delta_{ak}^{\textrm{\mbox{min}}}\leq\Delta_{ak}\leq\mbox{\ensuremath{\Delta}}_{\textrm{\mbox{max}}}^{\textrm{\mbox{F}}}$,
the transverse mode cannot exert a negative effect on the single-photon
transport. We point out that the features for $T$ and $P_{2L}$ versus the $b$ mode photon-atom
coupling $\gamma_{b}$ at the single-photon resonance remain similar with that in the linear
waveguide [Figs.~\ref{fig:T1}(b) and~\ref{fig:T1}(d)].


To show this more explicitly, how the transverse mode plays a role in the
photonic Feshbach resonance, we plot the photonic Feshbach resonance
peak position $\Delta_{a}^{\textrm{\mbox{F}}}$ in Fig.~\ref{fig:T_P2L}
versus the transverse mode coupling strength $\gamma_{b}$ in Fig.~\ref{fig:Detu_Fesh_gamb}. As the curve shows, $\Delta_{a}^{\textrm{\mbox{F}}}$
is nearly a linear curve and decreases when increasing the transverse-mode
coupling strength. This phenomenon agrees with the properties of $T$
shown in Fig.~\ref{fig:T_P2L}. Since $\Delta_{a}^{\textrm{\mbox{F}}}$
is also a component of the bound-state energy (\ref{eq:bs}) in the
transverse mode, except for a constant $\omega_{0}$, this curve also
shows the bound-state-energy dependence on the transverse-mode coupling
strength.

\begin{figure}
\begin{centering}
\includegraphics[bb=28bp 0bp 510bp 346bp,scale=0.55]{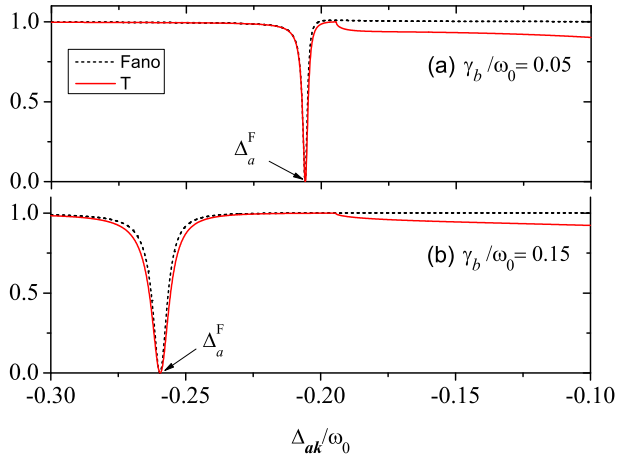}
\caption{\label{fig:Fano_T-1}(Color online) Comparison between the quasi-Fano
line shape (red solid line) for transmission coefficient T around
Feshbach resonance and the Fano line shape (black dashed line) given
by Eq. (\ref{eq:f}). (a) $q=10^{-4}$, $d=10^{-3}$; (b) $q=10^{-4}$,
$d=10^{-5/2}$. Other parameters are the same as in Fig. \ref{fig:T1}.}
\end{centering}
\end{figure}

We also find the line shape for the photonic Feshbach resonance is very close
to Fano line shape \cite{Fano1961PR}, which is compared with the Fano line in Fig. \ref{fig:Fano_T-1}
by defining the Fano function
\begin{equation}
f=\frac{\left(\Delta_{ak}-\Delta_{a}^{\textrm{\mbox{F}}}+q\right)^{2}}{\left(\Delta_{ak}-\Delta_{a}^{\textrm{\mbox{F}}}\right)^{2}+d^{2}}.\label{eq:f}
\end{equation}
We call our line shape quasi-Fano line. Here, we would like to point out that the similar resonance originated
from a bound state in higher transverse modes has also been discovered
in an electronic quasi-1D waveguide \cite{Levinson1993PRB,Stone1994PRB}.
The resonance line shape is Fano type for electrons in their results
\cite{Stone1994PRB}. However, compared with the resonance found in
Refs. \cite{Levinson1993PRB,Stone1994PRB} for electrons, the similar
resonance induced by the bound state for photons we find occurs exactly
at the bound-state energy, while the resonance position for electrons
acquires a shift in electronic waveguides~\cite{Levinson1993PRB,Stone1994PRB}.

Finally, we would like to estimate some parameters  for the conditions
when the additional transverse mode is involved for the study of single-photon
transport. Usually, we can ignore the influence of the $b$ mode,
when the $a$ mode is close to resonance of an atomic transition
while the $b$ mode is off-resonance. We now estimate the quantitative
condition by assuming that the effect of mode $a$ is 100 times
that of mode $b$. Namely, when
\begin{equation}
100\times\frac{g_{2}}{\left|\omega_{bk}-\omega_{0}\right|}\leq\frac{g_{1}}{\left|\omega_{ak}-\omega_{0}\right|}\label{eq:cri}
\end{equation}
or
\begin{equation}
L_{x}\leq c\frac{\left(\sqrt{2}-1\right)\pi}{\omega_{0}}\equiv L_{c},\label{eq:c}
\end{equation}
the transverse mode $b$ cannot affect the single-photon transport.
To obtain Eq.~(\ref{eq:c}), we have used $\left|\omega_{b}^{\textrm{\mbox{cut}}}-\omega_{a}^{\textrm{\mbox{cut}}}\right|\geq\left|\omega_{bk}-\omega_{ak}\right|$
and $g_{2}=g_{1}$. However, when the transverse size $L_{x}$ of
the waveguide is larger than the critical size $L_{c}$, $L_{x}>L_{c}$,
the transverse mode should be taken into account. For a 1D circuit
system with $\omega_{0}\simeq10\mbox{ GHz}$~\cite{Tsai2010Science},
then $L_{c}\simeq3.9\mbox{ c}\mbox{m}$. For a 3D optical cavity system
with $\omega_{0}\simeq2.21\times10^{6}\mbox{ GHz}$~\cite{Kimble2004Sci},
then $L_{c}\simeq176.6\mbox{ nm}$.

In addition, when the photon frequency $\sim10^{9}$ GHz, such as
x rays~\cite{Rohlsberger2010Sci,Rohlsberger2012Nat}, the TLS with
transition energy $14.4\mbox{ keV}$ (the nuclear transition of $^{57}F_{e}$),
corresponding to $\omega_{0}\simeq3.48\times10{}^{9}\mbox{ G}\mbox{Hz}$,
then the critical size becomes $L_{c}\simeq1.12\;{\AA}$. Experimentally,
this $1.12\;{\AA}$ looks too difficult. Therefore, it is very
necessary to consider the transverse-mode effect in the single-photon
transport in a waveguide with finite cross section. A finite-cross-section
waveguide is closer to our experimental quantum coherent device
design and fabrication. Taking advantage of the finite-cross-section
waveguide will ease the stringent requirements on realizing quantum-coherent
devices.

\section{Conclusions and discussions}

We studied the finite cross-sectional effect of the waveguide on single-photon
transport. To mainly characterize the finite cross-section effect
of the waveguide, we pick out one of the numerous transverse modes,
whose eigenfrequency is closest to that of the transport mode. We
consider the transport properties of a single photon in such a finite
cross-section waveguide by calculating the transmission, reflection
coefficients and the single-photon loss probability. By using a quadratic
dispersion relation, we find a bound state and two quasibound states
~\cite{Petrosky2007PRL,Petrosky2008PTP,Gong2008PRA} emerging in such
a waveguide with a finite cross section, which will not occur in the
usual linear waveguide. Moreover, when the input photon energy is
resonant with the bound-state energy in the transverse mode, the photon
will be completely reflected. This is the photonic Feshbach resonance.
In addition, the input photon is also completely reflected when the
input energy of it is at a single-photon resonance with the TLS or
at the cutoff frequency allowed by the approximated quadratic waveguide.
The photonic Feshbach resonance and the cutoff frequency resonance
phenomena do not occur in a linear waveguide even in an infinitely idealized
1D waveguide.

Furthermore, as a result of transverse-mode coupling, the photon will
be lost when the input energy is above the maximum bound-state energy
regulated by the coupling strength between the transverse mode and
the TLS. Therefore, only when the input energy is below this maximum bound-state
energy, the single photon can safely pass through or be completely
reflected by the TLS instead of lost in some other transverse
mode even though in a finite cross-section waveguide.


\begin{acknowledgments}
We would like to thank P. Zhang and D. Z. Xu for helpful discussions.
This work is supported by National Natural Science Foundation of China
under Grants No.~11121403, No.~10935010, and No.~11074261, and National 973 program
(Grant No.~2012CB922104). T.S. has been supported by the EU under the IP project AQUTE. F.N. acknowledges partial support from
the Army Research Office, JSPS-RFBR Grant No. 12-02-92100, Grant-in-Aid
for Scientific Research (S), MEXT Kakenhi on Quantum Cybernetics,
and the JSPS-FIRST program.
\end{acknowledgments}

\narrowtext

\end{document}